\documentclass[useAMS]{aa}
\usepackage{txfonts}
\usepackage{epsfig,amsmath,mathrsfs}
\bibpunct{(}{)}{;}{a}{}{,} 

\usepackage[usenames,dvipsnames]{color}

\usepackage{graphicx}
\usepackage[]{hyperref} 
\usepackage{fontawesome}
\usepackage{bm}
\usepackage{verbatim}
\usepackage{amssymb}
\usepackage{xcolor}
\usepackage{units}
\usepackage{tikz}
\usepackage[tikz]{bclogo}
\usepackage[framemethod=tikz]{mdframed}
\newcounter{infobox}[section]
\renewcommand{\theinfobox}{\thesection.\arabic{infobox}}

\definecolor{gray}{RGB}{180,180,180}

\definecolor{DarkRed}{RGB}{195,0,0} 
\hypersetup{colorlinks,%
            citecolor=blue,%
            filecolor=black,%
            linkcolor=DarkRed,%
            urlcolor=blue}

\usepackage{pgfornament}

\begin{document} 

\title{Galaxy formation from a timescale perspective}
\author{Peter Laursen\inst{1,2}}
\institute{Cosmic Dawn Center (DAWN).
  \email{pela@nbi.ku.dk}.
  \and
  Niels Bohr Institute, University of Copenhagen, Jagtvej 128, 2200 Copenhagen N, Denmark.
  }
\date{\today}
\abstract{Timescales in astronomy comprise the largest range of any scientific discipline.
In the construction of physical models, this circumstance may both be a blessing and a curse.
For example, galaxy evolution occurs on typical timescales of hundreds of millions of years, but is affected by atomic processes on sub-second timescales, posing a challenge in analytical and, in particular, in numerical models.
On the other hand, the vast dynamic range implies that we can often make meaningful predictions by simply comparing characteristic timescales of the physical processes involved.
This review, aimed primarily at non-astronomer scientists, attempts to highlight some occasions in the context of galaxy formation and evolution in which comparing timescales can shed light on astrophysical phenomena, as well as some of the challenges that may be encountered.
In particular we will explore the differences and similarities between theoretical predictions of dark matter halos, and the observed distribution of galaxies.
The review concludes with an account of the most recent observations with the James Webb Space Telescope, and how they purportedly seem to defy the timescales of the currently accepted concordance model of the structure and evolution of the Universe, the $\Lambda$CDM model.
}
\keywords{galaxies: kinematics and dynamics --- galaxies: formation --- cosmology: theory.}

\maketitle

\section{Introduction}
\label{sec:intro}

No scientific field deals with a wider span of timescales than astrophysics.
From explosion mechanisms in dying stars and oscillations of neutron stars, to cosmic structure formation and the age of the Universe itself, more than twenty orders of magnitude prevail.
Counting Big Bang physics as ``astrophysics'' doubles or even triples this amount\footnote{The age of the Universe is $\sim10^{61}$ Planck times!}.
However, while this enormous dynamic range can be challenging to model under one umbrella, it may also be an advantage: Because the evolution of a system in general happens at a rate determined by the longer timescale, modified by shorter timescales, many characteristic features of various astrophysical phenomena can be understood simply by comparing relevant physical timescales.

Whereas the short end of this range of timescales is in many cases observable, an inherent problem in astronomy is that many processes occur on timescales much longer than a human life, or even a human civilization that would be able to pass knowledge down through generations.
How, then, may we say anything about how stars form, how galaxies evolve, or which phases the Universe has gone through?

Several approaches exist to answering this question: Observationally, while we cannot wait to witness the evolution of any given galaxy --- which to us puny humans appears frozen in time --- we are lucky to live in a universe where the speed of light is finite, but fast
.
As we peer deeper into the cosmos, we probe earlier and earlier epochs.
Observing large samples of galaxies throughout cosmic history then allows us to study their properties in a statistical sense --- numerous defining properties such as their star formation rates, the build-up of heavy elements and dust, and changes in their morphology.

On the theoretical side, as in all other fields of science we make use of \emph{models}.
Models that predict the evolution of these various physical properties, while being (largely) consistent with already known physical properties.
Models can be analytical, numerical, or a mixture thereof; the so-called semi-analytical models.
In this chiefly theoretical review we will have a look at how timescales can be used to make meaningful predictions about the most beautiful structures in the Universe; the galaxies.

Ever since the revelation that the Milky Way does not comprise the whole Universe, but that we are surrounded by countless similar ``island universes'' \citep[][see also info box~\ref{info:debate} on p.~\pageref{info:debate}]{Hubble1925}, we have strived to comprehend the origin of these magnificent entities.
Which physical processes may lead to their formation and govern their evolution, while at the same time resulting in the astounding observed heterogeneity in their properties?

Progress in this captivating field is not only an exquisite interplay between the increasing power of telescopes and detectors, improved observational techniques, and an progressively better theoretical understanding of cosmology and the Universe in general; advances are also stimulated by breakthroughs in such diverse fields as particle physics, chemistry, and computer science.
No man is an island \citep{Jovi1990}, and indeed the same is true for fields of science --- interdisciplinarity is essential to progress!

\section{Galaxy formation and evolution}
\label{sec:galaxies}

A brief outline of the physical processes leading to a galaxy can be summarized as follows:

In the early, expanding Universe, sufficiently large overdensities are able to withstand and detach themselves from this expansion, turn over, and collapse.
With more than five times as much dark matter (DM) as baryonic matter (i.e.~gas), the dynamics are initially dominated by the former.
Eventually the collapsing cloud will \emph{virialize} --- astronomers' term for reaching a dynamical equilibrium --- and come to a halt.
Gas, which unlike the collisionless DM is able to cool and fragment on small scales, condenses in the center of more extended DM halos, with the densest regions further collapsing to nurture the fundamental building blocks of the galaxies: stars.

In the very center, supermassive black holes form which accrete mass, ejecting excess energy as so-called \emph{active galactic nuclei}.
Meanwhile, dying stars inject not only energy but also heavy elements (in astronomy, everything heavier than helium is collectively called ``metals'').
With time, the interstellar medium is therefore enriched with metals, roughly half of which condenses to dust.
Dust particles, in turn, in sufficiently dense environs (i.e.~in the proto-stellar disks) coalesce to larger grains, pebbles, rocks, planetesimals, and eventually planets which, for all we know, are a necessary condition for life.

Thermal and kinetic feedback from stars and active galactic nuclei drive strong winds which may exceed the galaxy's escape velocity, enriching also the intergalactic medium with metals.
Star formation typically declines after an initial starburst, but may be sustained by continuous accretion of new material from the intergalactic medium, while new starbursts may be initiated by collisions with other galaxies, a process known as \emph{merging}.
Galactic winds, merging, and gas depletion is also responsible for some galaxies ceasing to form new stars, the process known as \emph{quenching}.

Which processes dominate will determine the nature of the galaxy, in particular its color (governed mainly by the age and metallicity of the stellar population, and the amount of dust) and its morphology: will it end up as a disky spiral galaxy, a featureless elliptical galaxy, or something else?

\begin{figure}[!t]
\minipage[t!]{\dimexpr0.98\linewidth-2\fboxsep-2\fboxrule\relax}
\begin{bclogo}[
    couleur=gray!20,
    epBord=1,
    arrondi=0.1,
    logo=\bcinfo,
    marge=8,
    ombre=false, 
    couleurBord=gray!60,
    barre=line]
    { \ \textsf{Cosmological redshift}}
    \small{\textsf{Arguably, the most essential concept in astronomy is the \emph{redshift} of light, denoted by the letter $z$.
    Well-known to other physicists as \emph{Doppler shift}, the result of a relative motion between emitter and observer, in extragalactic astronomy there is a second effect, dominating on all but the most local scales, caused by the cosmological expansion of the Universe.\vspace{1mm}\\
    As light travels through expanding space, its wavelength increases by a factor $1+z$, equal to the ratio of the size of the Universe at the time of observation and emission, respectively.
    Because the Universe has always been expanding, the observed redshift becomes a measure of both the age of the Universe at emission, the current distance, the ``lookback time'', the observed volume, and several other useful aspects.
    We therefore simply use the term ``redshift'' to talk about all these quantities, collectively.
    For instance, we refer to galaxy seen 3.3 billion years after the Big Bang as ``\emph{a redshift 2 galaxy}'', we observe ``\emph{the high-$z$ Universe}'' (with disparate definitions among different researchers), or we speak of ``\emph{the evolution of galaxies with redshift}.''\vspace{1mm}\\
        As the Universe grows, the relative expansion of the Universe decreases.
        Hence, the time span between, say, $z=0$ (today) and $z=1$ is much longer than between $z=10$ and $z=11$, namely 8 billion years (Gyr) and 60 million years (Myr), respectively.
        Partly for the same reason, however, the timescales of evolution --- not only of galaxies but for all physical processes --- were shorter at earlier times.\vspace{1mm}\\
        The relationship between redshift and other quantities is shown in Fig.~\ref{fig:redshift}.
    }}
\label{info:redshift}
\end{bclogo}
     \endminipage
\end{figure}

\begin{figure}[!t]
    \begin{center}
        \includegraphics[width=0.98\linewidth]{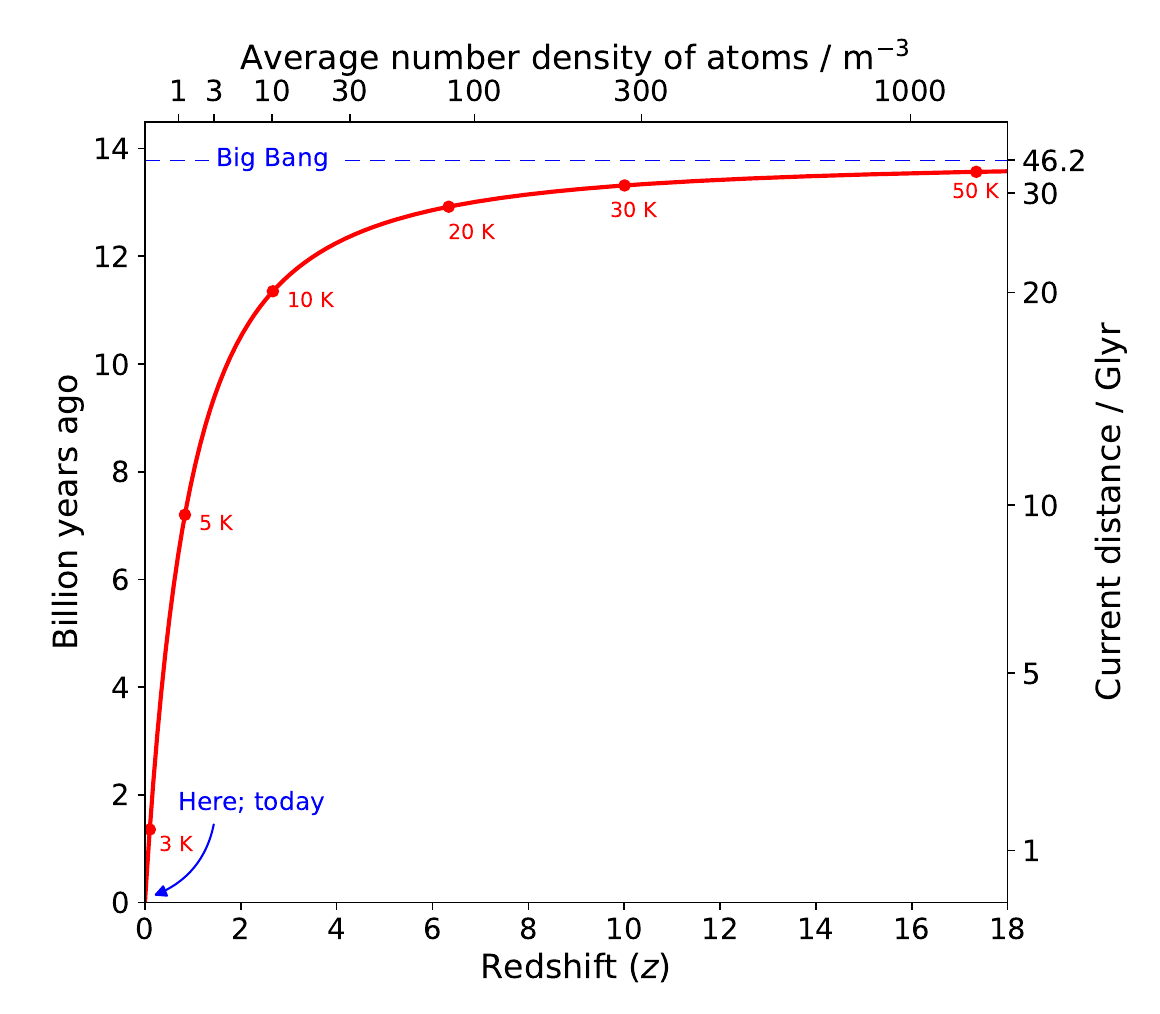}
        \caption{Relationship between the observed redshift of light emitted from a distant object (on the $x$ axis), and various properties of the Universe at the time the light was emitted:
        The left $y$ axis shows the ``lookback time'', i.e. how many years we look back in time; it ends at 13.8 Gyr, the age of the Universe.
        The right $y$ axis shows the current distance of the object; it doesn't merely follow the left axis Gyr$\rightarrow$Glyr, because the Universe expands while the light travels.
        The distance between us and the object at the time of emission was a factor $1+z$ smaller.
        The top $x$ axis shows the average density of atoms in the Universe, and the points along the plot line mark the minimum temperature of the Universe, set by the cosmic microwave background radiation.
        Because the relative expansion was larger in the past when the Universe was smaller, redshift increased more then, asymptotically approaching infinity for light emitted at the time of the Big Bang, 13.8 Gyr ago.}
        \label{fig:redshift}
    \end{center}
\end{figure}

\subsection{The luminosity function}
\label{sec:LF}

One of most popular ways of describing statistically a population of galaxies and its evolution through cosmic time is the \emph{luminosity function} (LF) --- the probability distribution function of the luminosities of galaxies in a given wavelength band, e.g. rest-frame ultraviolet light.
Here, ``rest-frame'' refers to the \emph{emitted} light, not the observed:
Due to cosmic expansion the radiation reaching us from distant sources is shifted to longer wavelengths (see Fig.~\ref{fig:redshift} and info box \ref{info:redshift} about the concept of redshift).
For instance, in the observer's frame UV radiation is redshifted into successively redder bandpasses, across the optical and out to the near-infrared for the most distant galaxies.
In order to compare luminosities of galaxies at different epochs, and hence different redshifts, we must first transform to the rest-frame.
Carefully taking into consideration observational pitfalls such as differential dust extinction and various selection biases (e.g.~faint galaxies are progressively more difficult to see at high redshifts/distances than bright galaxies), the physical properties and the evolution of a given galaxy type can then be studied through time.

A prime task of the field of cosmology is to describe the large-scale structure of the Universe:
How is matter distributed, and how does this distribution evolve.
Roughly 5/6 of matter is dark and not easily observed, so the best option we have is to look for the 1/6 that we \emph{can} see\footnote{There \emph{are} however ways to ``see'' the dark matter that do not rely on the light emitted from luminous matter associated with the dark matter: Gravitational lensing measures the distortion of background sources due to the foreground matter, thereby deducing its mass distribution.}.
But it is then crucial to determine how biased this luminous tracer of the total matter is.

All structure in the Universe is, ultimately, born out of primordial (quantum) fluctuations in the density field during the Big Bang, blown up to cosmological sizes during the process known as \emph{inflation}.
Observations of the cosmic microwave background\footnote{The relic radiation emitted 380,000 years after the Big Bang when the Universe had cooled enough for neutral hydrogen to form. The near-perfect Planckian spectrum emitted then has today been redshifted into the microwave regime.} show that, apparently, these fluctuations can be described by an almost scale-free power spectrum \citep{PlanckCollaboration2020}.
In other words, the Universe does not have a preferred scale, and neither does gravity (there are no characteristic masses in the Einstein equations).
We might therefore expect structure to form in a self-similar way: If the ratio between the number of structures with masses $M$ and $10M$ is $f$, then the ratio between the number of structures with masses $\frac{1}{10}M$ and $M$ is also $f$.
And if all halos contain the same baryonic fraction and are equally effective at converting gas to stars, we might expect the LF to reflect the distribution of masses in the Universe.

\begin{figure}[!t]
    \centering
    \includegraphics [width=0.48\textwidth] {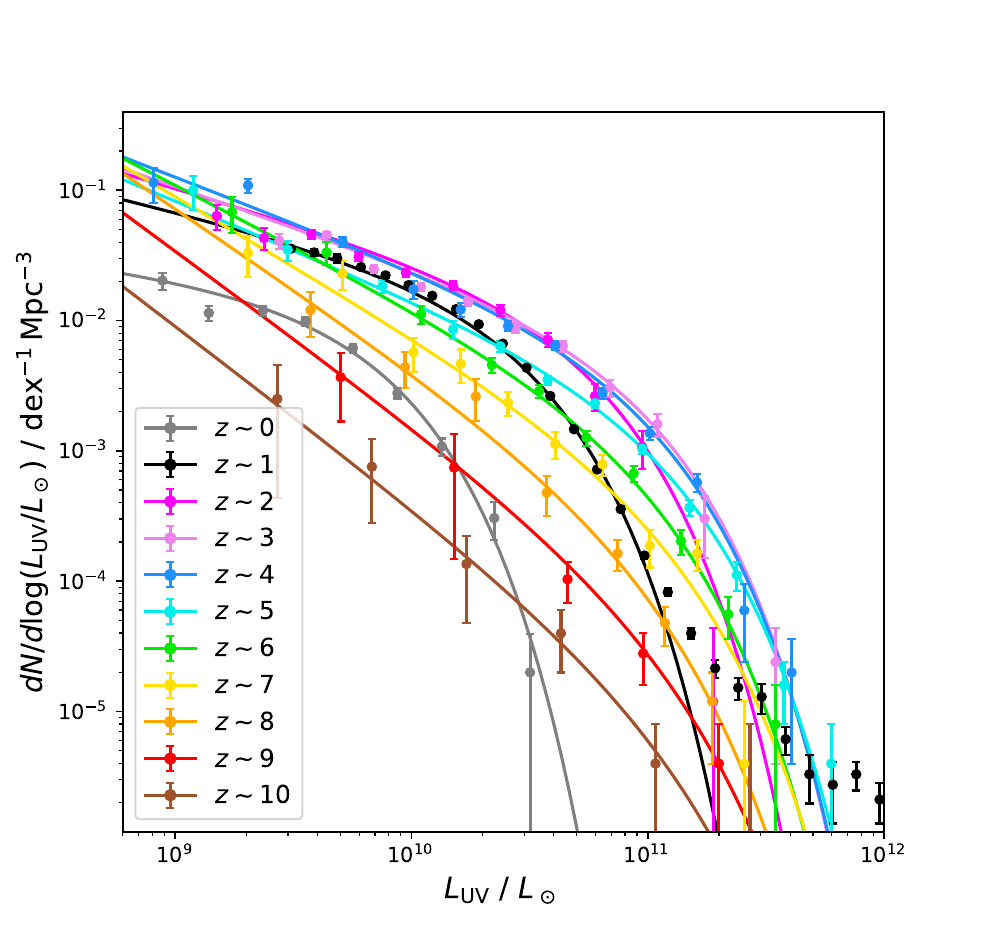}
    \caption{The evolving (ultraviolet) luminosity function of galaxies from the Universe was $\sim500\,\mathrm{Myr}$ ($z\sim10$) till today ($z\sim0$).
    Along the $y$ axis is differential number density, while the luminosity along the $x$ axis is given in Solar luminosities.
    The data are taken from
    \citet[][$z\sim0$]{Wyder2005},
    \citet[][$z\sim1$]{Moutard2020},
    \citet[][$z\sim2\text{--}9$]{Bouwens2021}, and
    \citet[][$z\sim10$]{Oesch2018}, and solid lines show the best-fit Schechter functions (Eq.~\ref{eq:LF}, but note that, due to the logarithmic binning, the faint-end slopes correspond to $\alpha+1$).
    In general, the number of (UV-emitting) galaxies is seen to rise until around $z\sim2$ (i.e. the overall amplitude increases), after which it declines again due both to decreased star formation and increased amounts of dust which absorbs light, in particular UV light.
    The excess at the high-mass end at $z\sim1$ is likely due to contribution from active galactic nuclei.}
    \label{fig:uvlf}
\end{figure}

The Universe, it turns out, is more complex than this, as will be elucidated further down.
Observationally, we do however see a LF that is remarkably universal --- that is, it can be rather accurately parametrized by same functional form for all galaxy types, and all redshifts:
As seen in Fig.~\ref{fig:uvlf}, at low luminosities the LF is characterized by a power law, while above a certain characteristic luminosity there is an exponential cut-off.
Observed luminosities are therefore often fitted with a so-called Schechter function \citep{Schechter1976} with three parameters that are a function of redshift: The faint-end slope $\alpha$, the normalization $\phi^*$, and the ``knee'' luminosity, $L^*$ (pronounced ``L-star''):
\begin{equation}
    \label{eq:LF}
    N(L)dL=\phi^* \left({\frac{L}{L^*}}\right)^{\alpha} e^{-L/L^*}{\frac{dL}{L^*}}.
\end{equation}
Typical estimates of the faint-end slope lie around $\alpha\simeq-1.5$ to $-1.1$ today, but is successively steeper at earlier times \citep[e.g.][]{Bouwens2021}.
The cut-off implies that there is in fact such a thing as a ``typical galaxy'', roughly comparable to our own Milky Way.
The fact that the number density of galaxies larger than the Milky Way declines exponentially fast means that, if one integrates the total stellar mass of all galaxies, Milky Way-sized galaxies dominate the total stellar budget in the Universe.
It is thus not entirely coincidental that we live in an $L^*$ galaxy, although other factors than just the sheer number of stars are important (not all galaxies are equally habitable; in particular the metallicity and the rate of supernovae may influence the probability of life forming and developing; see e.g.~\citealt{Dayal2015}).

The evolution of the LF with time has proved an invaluable tool for teaching us how galaxies have evolved since their formation.
For instance, the LFs depicted in Fig.~\ref{fig:uvlf} show that the number density of UV-emitting galaxies increased from the early Universe until around $z\sim2$, after which is declines.
The energetic UV radiation is emitted primarily by hot, massive stars with short lifetimes, and hence UV traces star-forming galaxies.
The decreasing amplitude of the UV LF reflects the fact that 1) the universal star formation rate density peaks at $z\simeq2$, after which it declines exponentially \citep{Madau2014}, and 2) metallicity and hence dust --- which absorbs preferentially shorter-wavelength light --- builds up over time.

Unlike earlier attempts to characterize galaxy populations, the Schechter function is not purely phenomenological, but has its roots in the underlying distribution of matter.
Through a variety of scaling relations, in certain wavelength regions the LF is expected to be related to the distribution of stellar masses --- the \emph{stellar mass function} (SMF).
The SMF, in turn, should be related to the distribution of DM halos.
In the following I will attempt to shed light on how from a timescale perspective we can understand, to a certain extent, the physics of the LF and the SMF, and their relation to the distribution of DM halos.

\subsection{The halo mass function}
\label{sec:HMF}

Galaxies reside in dark matter halos.
Despite our ignorance of the actual nature of DM, due to its collisionless nature its treatment is much simpler than when baryons are included.
One of the big discoveries of the 1970s was that, in contrast to the prevailing picture at the time \citep{Eggen1962}, galaxies do not form from huge, ``monolithically collapsing'' clouds that later fragment to stars.
Rather, it seems to be the other way round, with small structures forming first, later building up to larger structures in a hierarchical manner.
Thus dwarf galaxies are usually considered as early formed building blocks of larger systems, while halos on the scale of galaxy clusters are still currently assembling via infall of halos with their associated galaxies and galaxy groups.

This was first realized by \citet{Press1974}, and the theory was later refined by e.g. \citet{Sheth2002} and \citet{Tinker2008} who calculated the collapse of gravitating clouds from an initial smoothed density field.
Although the Press-Schechter formalism is a simple, analytical approach, its results are in qualitative agreement with later detailed modeling using $N$-body simulations\footnote{$N$-body simulations are the simplest form of numerical simulations of structure formation: They follow the trajectories of particles representing samples of matter, moving under the influence of gravity (see info box~\ref{info:adaptive} on p.~\pageref{info:adaptive}).}.
The resulting \emph{halo mass function} (HMF) is analogous to the LF and the SMF, but for DM halo masses rather than galaxy luminosities and stellar masses.
Like the LF, but for different reasons, its functional form is described by a power law with an exponential cut-off (see info box~\ref{info:obsuni}).

\begin{figure}[!t]
\minipage[t!]{\dimexpr0.98\linewidth-2\fboxsep-2\fboxrule\relax}
\begin{bclogo}[
    couleur=gray!20,
    epBord=1,
    arrondi=0.1,
    logo=\bcinfo,
    marge=8,
    ombre=false, 
    couleurBord=gray!60,
    barre=line]
    { \ \textsf{The observable Universe and the Hubble timescale}}
    \small{\textsf{
The exponential departure of the halo mass function from a universal power law derives from the Gaussian (or at least near-Gaussian) distribution of clumps in the primeval density field.
Structure forms on all scales, but because no information can be gained from outside the cosmic event horizon --- i.e.~the ``boundary'' marking the sphere within which light has had the time to travel since the Big Bang, the so-called ``observable Universe'' --- small structures form before larger ones.\vspace{1mm}\\
The relevant timescale that sets the maximum size of structures in the Universe at a given time is the \emph{Hubble time}, defined at a time $t$ as $t_\mathrm{H}(t) = 1/H(t)$, where $H(t)$ is the Hubble parameter that describes the expansion rate of the Universe.\vspace{1mm}\\
The expansion rate measures the recession velocity at a given distance, typically in km/s per ``mega-parsec'', or Mpc, where 1 parsec $\simeq$ 3.3 lightyears.\vspace{1mm}\\
The current value of $H(t)$, $H_0$, is called the Hubble constant and is roughly equal to $H_0\simeq70\,\mathrm{km}\,\mathrm{s}^{-1}\,\mathrm{Mpc}^{-1}$.
If the Universe had always expanded at its current rate, $t_\mathrm{H}$ would be equal to the age of the Universe.\vspace{1mm}\\
In most cases we can ignore physical processes that happen on timescales longer than $t_\mathrm{H}$.
For instance, this timescale sets the primordial abundances of elements created through nucleosynthesis in the first 20 minutes after the Big Bang.
    }}
\label{info:obsuni}
\end{bclogo}
     \endminipage
\end{figure}

Since baryons and DM was originally mixed in the mass ratio $\sim$1:5, one might expect that the LF would mimic the HMF, albeit shifted to lower masses.
But this is not at all what we see! Although they are both characterized by a power law and an exponential cut-off, the HMF cut-off is at much high masses than the halos hosting $L^*$ galaxies.
At lower and higher luminosities than $L^*$, the slope of the LF is significantly shallower and steeper, respectively, than the HMF.
Evidently, there must be some physical mechanism(s) suppressing the conversion of gas to stars in both small and large mass halos.

\subsection{Gas cooling in dark matter halos}
\label{sec:cooling}

How may we understand the shape of the LF? The first clues to this problem were provided by \citet{Rees1977} who simply compared some relevant timescales: 

A cloud of gas and DM that meets the Jeans criterion and is able to withstand the expansion of the Universe collapses first in a free fall.
In the simple case of a spherically symmetric collapse, it follows from the law of gravity that all particles of a pressureless gas meet in the center on the free-fall timescale $t_\mathrm{ff}$ given by
\begin{eqnarray}
    \label{eq:tff}
    t_\mathrm{ff} & = & \left( \frac{3\pi}{32 G \rho_\mathrm{m}} \right)^{1/2}
                    = \left( \frac{3\pi f_\mathrm{b}}{32 G n \mu m_\mathrm{H}} \right)^{1/2}\\
                  & \simeq & 0.8\,\mathrm{Gyr}\,\times\,\left(\frac{n}{10^{-3}\,\mathrm{cm}^{-3}}\right)^{-1/2},
\end{eqnarray}
where
$G$ is the gravitational constant and $\rho_\mathrm{m}$ is the mass density.
In the second step we have expressed this density in terms of the gas through the universal baryonic fraction $f_\mathrm{b} = \rho_\mathrm{b}/\rho_\mathrm{m} = 0.157$ \citep{PlanckCollaboration2020}, the particle density $n$, and the mean molecular mass\footnote{For a primordial gas of hydrogen and helium in the mass ratio $\frac{3}{4}$:$\frac{1}{4}$, the mean molecular mass is 1.23 or 0.59 if the gas is neutral or fully ionized, respectively.} $\mu$ in terms of the hydrogen mass $m_\mathrm{H}$.
In the early Universe, before dark energy played any role (see info box~\ref{info:expansion}), the density simply evolved as $\rho_\mathrm{m}(z) = \rho_\mathrm{m,0} (1+z)^3$ where $\rho_\mathrm{m,0} = \rho_\mathrm{DM,0} + \rho_\mathrm{b,0}$ is the present-day density.
Evaluating Eq.~\ref{eq:tff}, the free-fall timescale is seen to be of the order of (hundreds of) Myr at high redshifts, increasing to Gyr at low $z$.

\begin{figure}[!t]
\minipage[t!]{\dimexpr0.98\linewidth-2\fboxsep-2\fboxrule\relax}
\begin{bclogo}[
    couleur=gray!20,
    epBord=1,
    arrondi=0.1,
    logo=\bcinfo,
    marge=8,
    ombre=false, 
    couleurBord=gray!60,
    barre=line]
    { \ \textsf{Expansion of the Universe}}
    \small{\textsf{
        The Universe consists of several different components which affect its dynamics differently:
        Dark energy (DE) has a constant energy density; more DE is therefore created along with the expansion of the Universe.
        Matter (m) consists of dark matter (DM) and baryons (b) which both act ``normally''; its density decreases proportionally to the volume because the total amount is largely conserved. Because redshift is inversely proportional to the ``size'' of the Universe, or \emph{scale factor} $a$ --- i.e. the redshift of an object observed when the size of the Universe was $a$ compared to today is given by $1+z=1/a$ --- the density of matter evolves with redshift as
        $\rho_\mathrm{m}(z) = \rho_\mathrm{m,0} (1+z)^3$.
        Radiation (r) may of course be created and destroyed, but the number of photons is dominated by photons from the cosmic microwave background which are largely conserved. However, in addition its energy density is diluted due to its redshift, and the energy density therefore goes as
        $\rho_\mathrm{r}(z) = \rho_\mathrm{r,0} (1+z)^4$.\vspace{1mm}\\
        Because of the differing exponents, the different components dominate on different timescales:
        In the very early Universe, dynamics were dominated by radiation until an age of roughly $t = 50\,000\,\mathrm{yr}$, after which we entered the matter-dominated era.
        The future will bring an exponential expansion because DE took over at around $t\sim10\,\mathrm{Gyr}$, but it is a rather remarkable fact that we live in an era where both matter and DE governs the expansion.
    }}
\label{info:expansion}
\end{bclogo}
     \endminipage
\end{figure}

But the gas is \emph{not} pressureless, and an initially adiabatic collapse therefore increases the gas temperature until thermal pressure prevents the cloud from contracting further and the gas is shock-heated to the \emph{virial temperature}:
\begin{eqnarray}
    \label{eq:Tvir}
    \nonumber
    T_\mathrm{vir} & = & \frac{\mu m_\mathrm{H}}{2 k_\mathrm{B}} \frac{G M_\mathrm{vir}}{R_\mathrm{vir}}\\
    & \sim & 2\times10^4\,\mathrm{K}\,\times\, \left( \frac{M_\mathrm{vir}}{10^8M_\odot} \right)^{2/3}
          \left( \frac{1+z}{10} \right),
\end{eqnarray}
which is found by equating the thermal energy of particles to the potential energy of a gravitating sphere.
Here $k_\mathrm{B}$ is the Boltzmann constant, and $M_\mathrm{vir}$ and $R_\mathrm{vir}$ are the virial mass and radius of the cloud, respectively (i.e.~the subscript ``vir'' is used to denote values that the cloud reaches once it has virialized).
In the second line we have used that a collapsing cloud reaches an overdensity with respect to the average matter density $\bar{\rho}_\mathrm{m}$ of $\Delta\sim180$ \citep[e.g.][]{Binney2008}, as well as the fact that $\bar{\rho}_\mathrm{m}$ increases with redshift as $(1+z)^3$.

In order to form stars, the gas needs to be much colder, $T\sim10^2\,\mathrm{K}$, than implied by Eq.~\ref{eq:Tvir} for a typical halo of $M_\mathrm{vir}\sim10^{12} M_\odot$.
In other words, only the smallest halos would be able to form any stars at all.
The key to more massive halos being able to host stars is the gas' ability to \emph{cool} through radiative processes, that is, they can convert their kinetic energy to radiation which is then able to leave the system.

Various physical mechanisms dominate the cooling process at different temperatures.
At very high temperatures, $T\sim10^{6\text{--}7}\,\mathrm{K}$ and above, where everything is ionized, cooling takes place through free-free emission, or Bremsstrahlung, where accelerated charged particles (mostly electrons) emit photons.
If the gas is metal-rich, many electronic transitions at various wavelengths exist that may cool the gas efficiently.
But in the early Universe, before the gas was polluted with metals, the only efficient cooling mechanisms at lower temperatures was through hydrogen and helium:
At low temperatures, $T\sim10^4\,\mathrm{K}$, collisions with free electrons excite the atoms, with subsequent de-excitation leading to emission.
At slightly higher temperatures, $T\sim10^{4\text{--}5}\,\mathrm{K}$ , most atoms are ionized; subsequent recombinations that do not go directly to the ground state (which would just result in a new ionizing photon) cascade down through intermediate states, emitting low-energy photons.
The effectiveness of these pristine elements peaks at $T\sim2\times10^4\,\mathrm{K}$ and $T\sim10^5\,\mathrm{K}$ for hydrogen and helium, respectively \citep[see e.g.][]{Katz1996}.

The total cooling function resulting from these processes is denoted $\Lambda(T)$ and can be calculated from quantum physics \citep[e.g.][]{Sutherland1993}; it is defined such that the cooling \emph{rate} is
\begin{equation}
    \label{eq:Lambda}
    \frac{dE_\mathrm{cool}}{dt} = n_\mathrm{H}^2 \Lambda(T),
\end{equation}
where $n_\mathrm{H}$ is the number density of hydrogen.
The timescale for the cloud to radiate away its kinetic energy $K = \frac{3}{2}n k_\mathrm{B}T$ is thus
\begin{eqnarray}
    \label{eq:tcool}
    t_\mathrm{cool} & = & \frac{K}{dE_\mathrm{cool}/dt}
    = \frac{3^5}{2^5}\frac{k_\mathrm{B}T}{n\Lambda(T)}\\
    & \simeq & 3.3\,\mathrm{Gyr}\times
    \left(\frac{T}{10^6\,\mathrm{K}}\right)
    \left(\frac{n}{10^{-3}\,\mathrm{cm}^{-3}}\right)^{-1}\!
    \left(\frac{\Lambda(T)}{10^{-23}\,\mathrm{erg}\,\mathrm{s}^{-1}\,\mathrm{cm}^3}\right)^{-1},
\end{eqnarray}
where we have used that $n = (9/4)n_\mathrm{H}$ for a primordial, ionized gas.

\begin{figure*}[!t]
    \centering
    \includegraphics [width=0.75\textwidth] {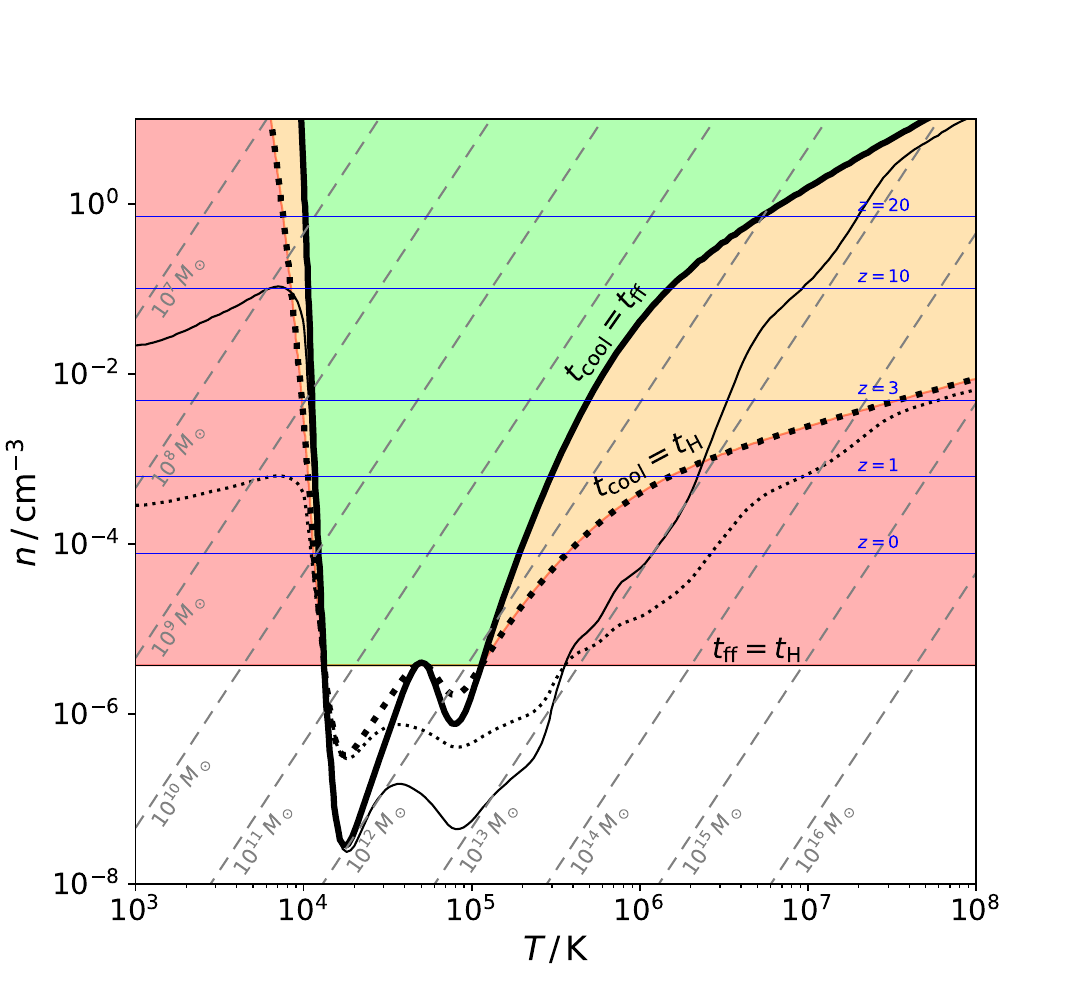}
    \caption{Particle density $n$ vs.~gas temperature $T$ of gravitating gas clouds.
    Below the horizontal line at $n\sim3\times10^{-6}\,\mathrm{cm}^{-3}$, densities are so low that clouds still expand with the Universe.
    The thick, black loci mark the $\{T,n\}$ values where the cooling timescale $t_\mathrm{cool}$ equals the dynamical timescale $t_\mathrm{ff}$ (solid) and the Hubble time scale $t_\mathrm{H}$ (dotted), for a primordial, metal-free gas.
    The similar thin, black lines show the same, but for Solar metallicity.
    Clouds in the red domain have virialized and are in hydrostatic equilibrium, but their cooling time is longer than the age of the Universe.
    A cloud in the orange domain is in a quasi-static equilibrium, slowly contracting.
    As time goes, it may enter the green domain where pressure is unable to support the cloud, and it will collapse on the free-fall timescale.
    Horizontal, blue lines mark typical densities of virialized halos at the indicated redshifts (with overdensities of $\Delta\sim180$).
    Slanted, gray dashed lines show $\{T,n\}$ values of halos with the indicated virial masses (Eq.\ref{eq:Tvir}).
    }
    \label{fig:cooling}
\end{figure*}

From Eqs.~\ref{eq:tff} and \ref{eq:tcool} we see that the question of whether or not a cloud will be able to form stars efficiently boils down to an interplay between its density and its temperature.
Figure~\ref{fig:cooling} compares the cooling timescale to the free-fall timescale in an $n$ vs. $T$ plane. 
The plane unveils three distinct domains in which a virialized gas cloud may find itself.
These domains are separated by two loci in thick solid and dotted lines that trace out, respectively, the sets of $\{T,n\}$ where the cooling timescale $t_\mathrm{cool}$ (Eq.~\ref{eq:tcool}) equals the free-fall timescale $t_\mathrm{ff}$ (Eq.~\ref{eq:tff}) and the Hubble time $t_\mathrm{H}$ (which is essentially the age of the Universe).
The loci have been calculated using a primordial, zero-metallicity cooling function (the $\Lambda$ in Eq.~\ref{eq:Lambda}).
The corresponding curves for a Solar metallicity, characteristic of the present-day Universe, are shown with thin lines.
Both cooling functions have been adopted from \citet{Smith2008}.
The high-metallicity loci are lower in the plane because metals facilitate cooling, so smaller densities are needed.
At $T\sim10^5\,\mathrm{K}$ cooling is accelerated in particular by carbon and oxygen, while at $T\sim10^6\,\mathrm{K}$ the many transitions of iron dominate.
Below $T\sim10^4\,\mathrm{K}$, there are virtually no free electrons and hence no cooling for a metal-free gas, but as metallicity increases molecular cooling comes into play (mostly H$_2$ and CO).

The domains are colored according to dominating timescales:
In the green domain, the cooling timescale is less than the free-fall timescale, so dissipative processes dominate over dynamical processes, and the gas collapses more or less freely with negligible thermal pressure.
As the density increases, cooling happens faster and faster, and eventually the densest region will be able to form stars.
In other words: Halos in the green domain can form galaxies.
Clouds in the orange domain cool slowly and contract, but maintain a (quasi-)hydrostatic equilibrium. 
As they contract, they may cross the $t_\mathrm{cool}=t_\mathrm{ff}$ locus and collapse, or they may be ``overtaken'' by the locus evolving with time and hence metallicity (i.e. the green domain eventually expands into the orange domain).
In the red domain the cooling timescale is longer than the Hubble time, so clouds are still in hydrostatic equilibrium.
Finally, clouds below the horizontal $t_\mathrm{ff}=t_\mathrm{H}$ line are not dense enough to have virialized yet. From Eq.~\ref{eq:tff}, this corresponds to a density
\begin{equation}
    \label{eq:n_ffH}
    n = \frac{3\pi f_\mathrm{b}}{32 G \mu m_\mathrm{H} t_\mathrm{H}^2}
      \simeq 3\times10^{-6}\,\mathrm{cm}^{-3}.
\end{equation}


The slanted lines show densities and temperatures of halos of constant mass. We see that clouds of $M_\mathrm{vir}\sim10^{11}\text{--}10^{12}\,M_\odot$ can cool, collapse, and start forming stars even at low densities while at higher and lower masses, cooling is less efficient.
In particular halos above $M_\mathrm{vir}\sim10^{12}\,M_\odot$ ($10^{13}\,M_\odot$) should not be able to form any galaxies at all from a pristine (Solar metallicity) gas.
Low-mass halos, however, should be able to cool and form galaxies at sufficiently high densities.
This simple argument explains why small galaxies are able to form in the early Universe, and why galaxy clusters in the present-day Universe, which have typical halo masses of $M_\mathrm{vir}\sim10^{14}\text{--}10^{15}\,M_\odot$, still contain large amounts of hot gas.

To investigate the physics governing galaxy formation we wish to compare the SMF to the HMF.
We don't observe stellar masses, however.
We observe luminosities, and then in some way interpret this in terms of stellar masses.
Converting from an observed LF to an inferred SMF is not straightforward, because the amount of light detected from a given amount of stars depends on many factors such the age of the stellar population, the metallicity, and interstellar dust.
In particular the short wavelengths are susceptible to dust, but also to age because the most massive and blue stars die out fast, leaving the red stars.
Therefore SMFs in infrared wavelengths are typically more robust than in the UV.
A discussion of the conversion from LFs to SMFs is beyond the scope of the current text, but can be found e.g.~in \citet{Song2016}.

\subsection{The overcooling problem}
\label{sec:overcooling}

\citet{Rees1977} and coeval authors \citep[e.g.][]{Binney1977,Silk1977} used this ``selective cooling'' to explain the exponential cut-off in the galaxy LF, and why we should not expect the SMF to simply mirror the HMF, shifted by the baryon fraction $f_\mathrm{b}$.
In the calculations of the range of efficient cooling in Fig.~\ref{fig:cooling} we assumed a single density.
However, this was the \emph{average} density inside the virial radius; in reality the halo will have a radially decreasing density profile with much higher densities in the center, so at least some of the gas should be able to cool.

Taking into account the fact that gas accumulates and condensates in the bottom of the potential well created by the halo, \citet{White1978} calculated a LF.
The result was still that the typical galaxy is hosted by a $M_\mathrm{vir}\sim10^{12}\,M_\odot$ halo.
But now astronomers faced another nuisance:
The high density allowed galaxies to cool \emph{too} efficiently, vastly overpredicting the number density of especially small galaxies.

The rescue for this ``overcooling'' problem is believed be, mainly, three physical processes known as \emph{photoionization}, \emph{merging}, and \emph{feedback}.

\subsection{Photoionization}
\label{sec:reionization}

In the ``dark ages'', before the advent of the first luminous sources, the Universe was filled with neutral gas which absorbed all light blueward of the ionization threshold at $912\,{\AA}$ \citep{Gnedin2000,Barkana2001}.
Strong UV radiation created ionized bubbles around the first galaxies which percolated the Universe and eventually overlapped.
The resulting meta-galactic UV background permeated and heated the intergalactic medium and increased its pressure, preventing accretion of gas onto halos, as well as reduced the rate of radiative cooling within the halos \citep{Benson2002}, inhibiting star formation.
Small halos are particular vulnerable to this effect, resulting in a shallower slope in the LF.
The UV background increased with time, peaking around $z\sim2$ and then decreased, with a characteristic value of $\Gamma\sim10^{-12}\,\mathrm{photons}\,\mathrm{s}^{-1}$, leading to a characteristic timescale for photoionization of $t_\mathrm{ph} \sim 1/\Gamma \sim 3\times10^4\,\mathrm{yr}$; much shorter than the typical dynamical timescales involved.
This peak in the UV background coincides roughly with the cosmic peak of star formation, but is not directly related: Ionizing photons from galaxies from galaxies peak somewhat earlier, around $z\sim4$, but are eventually outshone by quasars by a factor of a few \citep[]{Haardt2012,Khaire2019}.

Although various observational probes constrain this so-called \emph{epoch of reionization} to take place roughly when the Universe was 400--900 Myr old \citep[e.g.][]{Mason2019}, the exact timescales involved are still not entirely clear:
What was its duration (when did it start and end; was it short and intense or more prolonged?), which sources contributed more (small galaxies, massive galaxies, or even quasars?),
and what was its topology (small and fizzly vs. large bubbles)?

Analytical solutions exist for calculating the propagation of an ionized bubble around a source in a homogeneous gas, relating the output rate of ionizing photons to the timescale of recombination \citep{Stromgren1939,Dopita2003}.
Comparing these timescales, however, only puts weak constraints on the epoch of reionization; the largest advances in this field have been made through numerical simulations which can capture the highly non-linear nature of the involved densities, temperatures, velocities, etc.~\citep[see e.g.][for recent advances regarding the topology of reionization]{Hutter2020,Perez2022}.
A paramount problem in such calculations is the enormous range in spatial scales involved (see info box~\ref{info:adaptive}):
Typical bubbles have sizes of the order of ten(s of) Mpc \citep{Wyithe2004,Giri2018}, but photons begin ionizing gas already on sub-parsec scales, in the molecular clouds enshrouding the stars.
Most simulations have therefore had to focus either on the escape of ionizing radiation through the interstellar medium of individual galaxies \citep[e.g.][]{Haid2019}, or on the ionization of the large scales of the intergalactic medium, treating individual galaxies simply as (unresolved) photon sources \citep[e.g.][]{Jensen2013,Jensen2014}, although intermediate-scale simulations also do exist \citep[e.g.][]{Rosdahl2022}.

\begin{figure}[!t]
\minipage[t!]{\dimexpr0.98\linewidth-2\fboxsep-2\fboxrule\relax}
\begin{bclogo}[
    couleur=gray!20,
    epBord=1,
    arrondi=0.1,
    logo=\bcinfo,
    marge=8,
    ombre=false, 
    couleurBord=gray!60,
    barre=line]
    { \ \textsf{Dealing numerically with a multiplicity of timescales}}
    \small{\textsf{A fundamental challenge in numerical simulations --- not only in astrophysics --- is the vast dynamic range in both space and time.
    Computers are limited by both memory and processing power, and a trade-off must necessarily be made between simulating sufficiently large spatial and temporal scales, and resolving these scales sufficiently to capture the small-scale physics.
    Various approaches exist to alleviating these limitations; cosmological simulations, modeling a more or less representative chunk of the Universe, usually imploy an adaptive resolution.\vspace{1mm}\\
    Structures can be simulated using \emph{particles} that each represent a given amount of mass (gas, dark matter, and/or stars).
    This is analogous to simple $N$-body codes, but particles are ``smoothed'' in space \citep{Gingold1977,Lucy1977}.
    The particles follow trajectories dictated by forces (gravitational, hydrodynamical, magnetic, etc.), and resolution thus increases automatically in high-density regions.
    Alternatively, structure can be simulated on a fixed grid of cells, analogous to pixels in a picture, but in 3D.
    Adaptive resolution can then be achieved by refining cells according to a pre-defined criterion (usually a density threshold), recursively splitting cells up in 8 cells \citep{Berger1984,Berger1989}.
    The two schemes both have their pros and cons.
    Additionally, a relatively new technique is a ``moving mesh'' that incorporates the best of two worlds; here space is split up into cells, but the cells have arbitrary (polyhedral) shapes, and they are not fixed in space but move around \citep{Springel2010}.\vspace{1mm}\\
    Not only space, but also time may be refined.
    In dense regions where matter undegoes large accelerations, numerical errors will be introduced if the time step is too big.
    In order to avoid wasting time \emph{when} matter is not dense, time steps may adapt to the density.
    Likewise, to avoid wasting time \emph{where} matter is not dense, different spatial domains may use different time steps such that the equations of motion are advanced first on the smallest time steps, and then on successively larger scales until the largest time step ``catches up'' with the smaller.
    }}
\label{info:adaptive}
\end{bclogo}
     \endminipage
\end{figure}

\subsection{Galaxy merging}
\label{sec:merging}

Even today, the separation between galaxies is not vastly larger than their cross section, and at $z\sim10$, where number densities were $10^3$ times larger, encounters between galaxies were frequent\footnote{For instance, the distance between the Milky Way and Andromeda is 2.5 million lightyears, some 20$\times$ the size of their gaseous disks, and only a few times larger than the extent of the DM halos.}.
Galaxies therefore interact gravitationally and occasionally merge with other galaxies.
Minor mergers, where a galaxy of mass $M_1$ accretes a smaller galaxy of mass $M_2 \lesssim M_1/10$ are frequent at all times, but even major merger events, with $M_1\sim M_2$, happen on average once per Hubble time \citep{Lacey1993}.
The effect of this process is to increase the number of large galaxies, at the expense of small galaxies \citep[e.g.][]{Cole1991,Kauffmann1993}, effectively tilting the LF anti-clockwise.

To first order, the relevant timescale for accretion of satellite galaxies is set by a physical process known as \emph{dynamical friction}, described first by \citet{Chandrasekhar1943}, where the gravity of a large body moving through an atmosphere of collisionless gas (in this case the DM halo) creates a ``wake'' of increased mass, exterting a drag on the body.
For minor mergers, the timescale with which the orbit of a satellite decays decreases with increasing mass; \citet{Cole1994} found that a simple model with
\begin{equation}
    \label{eq:tmerge}
    t_\mathrm{merge} \sim t_\mathrm{dyn} \left( \frac{M_1}{M_2} \right)^{0.25},
\end{equation}
where the dynamical timescale $t_\mathrm{dyn}$ is essentially equal to the free-fall timescale, follows the trend of numerically simulated mergers.
For sufficiently large ratios between $M_1$ and $M_2$
the merging timescale exceeds the Hubble time, consistent with the fact that massive galaxies still have satellites today.
More sophisticated models \citep[e.g][]{Boylan-Kolchin2008}
yield similar conclusions.

\subsection{Feedback}
\label{sec:feedback}

Even with the physical mechanisms described above, galaxy formation models struggled to match observed LFs, apparently producing too many stars.
This annoyance entailed the need for \emph{feedback} processes which could regulate star formation.
Early models \citep[e.g.][]{White1991,Somerville1999,Efstathiou2000} employed only supernova feedback which inject both thermal energy and momentum into the interstellar medium. 
That this process is able to affect star formation can be understood first from a simple timescale perspective:
Star formation in galaxies occurs basically on the local dynamical timescale which is of the order 100 Myr. In contrast, supernovae result from the explosions of the most massive stars which has life times of $\lesssim 10$ Myr.

Supernova feedback not only reheats the cold gas but may also expel it altogether in \emph{galactic winds}, and is thus very efficient at inhibiting star formation. The effect is particular pronounced in low-mass galaxies which have a shallower gravitational potential, allowing gas to escape more easily \citep{Dekel1986,Kauffmann1993,Natarajan1999}.
In fact, it turned out to be a bit too efficient, because the expelled gas then becomes available for accretion onto more massive halos (galaxy groups and clusters) where at late times it cools and forms galaxies much more massive than observed \citep{Benson2003}.

Various mechanisms were proposed to solve this problem of too many massive galaxies.
Arguably the most promising mechanism is feedback from active galactic nucleir, i.e. gas accreting onto supermassive black holes in the center of galaxies, radiating excess energy away at extreme energies.
This effect has been successful at matching LFs at low and high redshifts \citep[e.g.][]{Croton2006,Bower2006,Somerville2008,Mccarthy2011}.

\subsection{Cold accretion}
\label{sec:coldmode}

During the latest two decades it has become clear that another channel should exist for galaxy formation, namely the so-called ``cold mode accretion''.
Whereas gas in the previously discussed ``hot mode'' was assumed to accrete isotropically, hydrodynamical simulations revealed that gas may flow along dense filaments and reach the halo center without being shock-heated \citep{Birnboim2003}.
At low redshifts, this mechanism is most efficient for halos of $M_\mathrm{vir}<10^{12}\,M_\odot$, but becomes efficient for increasingly larger masses for $z\gtrsim2$ \citep{Dekel2006}.
Observational evidence for cold accretion is sparse, but because a large fraction of the gas is never heated above $T\sim2\times10^4\,\mathrm{K}$ \citep[e.g.][]{Keres2005}, cooling should happen primarily via collisional excitation of hydrogen to its first excited state, leading to emission of Lyman $\alpha$ radiation at 1216 {\AA} \citep[see e.g.][for recent observational evidence]{Daddi2021,Daddi2022}.

\subsection{From halo mass to stellar mass}
\label{sec:hmf2smf}

\begin{figure*}[!t]
    \centering
    \includegraphics [width=0.75\textwidth] {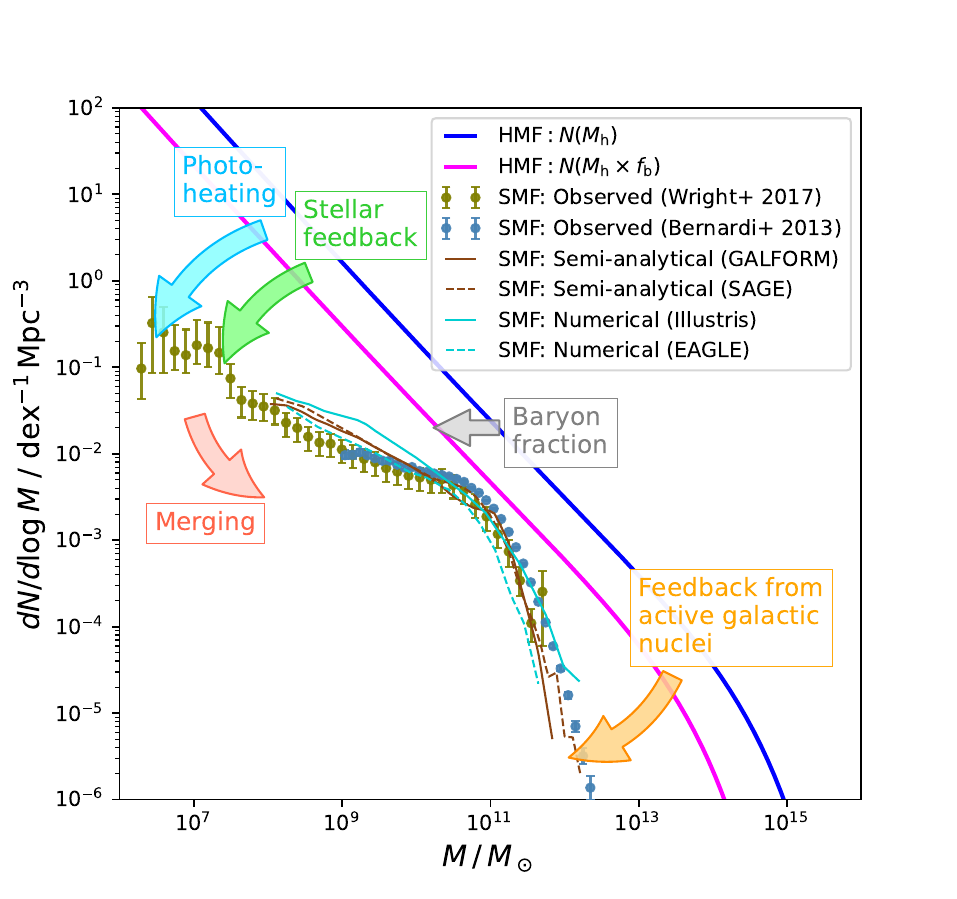}
    \caption{Number density of halo masses and stellar masses as a function of mass.
    Theoretical predictions from a halo mass function are shown in blue and magenta for total halo masses and corresponding baryon masses, respectively.
    Observed stellar mass functions are shown at the low-mass end (olive points) and the high-mass end (steel blue points).
    Photoionization and stellar feedback is believed to suppress the mass function at the low-mass end (blue and green arrows), while active galactic nucleus activity suppresses the high-mass end (orange arrow).
    Marging (red arrow) also decreases the number of small halos, thereby increasing the mass of larger halos.
    These mechanisms are incorporated in semi-analytical (brown) and numerical (cyan) models, which are seen to provide a reasonable match to the observations.}
    \label{fig:hmf2smf}
\end{figure*}

The effect of the physical processes described in Secs.~\ref{sec:LF}--\ref{sec:coldmode} is shown in Fig.~\ref{fig:hmf2smf}.
The blue line shows the HMF at $z=0.1$ (``almost today'').
It is a \citet{Sheth2002} HMF with a $z$-dependent correction factor which brings the analytical prediction very close to results from a large-scale, $\sim8$ billion particle, cosmological simulation \citep{Klypin2011};
its functional form is described in \citet{Laursen2019}.
Multiplying masses by the cosmic baryon fraction $f_\mathrm{b}$ shifts the HMF to the left, indicated by the black arrow.
If galaxies at all masses were able to cool and convert all their gas to stars, observed SMFs would look like the magenta line.
Reality is quite different: Observed SMFs at $z\sim0.1$ at the low-mass end \citep{Wright2017} and the high-mass end \citep{Bernardi2013} are shown with olive green and steel blue points, respectively.

The physical mechanisms thought to be responsible for this drastic suppression are indicated by the arrows.
The observations are compared to four different models that include these mechanisms: 
The brown lines show two semi-analytical models, GALFORM (\citealt{Gonzalez-perez2014}, but taken from \citealt{Somerville2015}) and SAGE \citep{Croton2016}, while the cyan lines show results from numerical, hydrodynamical simulations, Illustris \citep{Vogelsberger2014} and EAGLE \citep{Schaye2015}.
While they are not perfect fits, all models are seen to capture rather well the shape of the SMFs.
A breakdown of the relative impact of the individual processes can be found in e.g. \citet{Benson2003} and \citet{Bower2012}.

The preceding discussion has illucidated how simple timescale arguments may both be helpful and provide insight into complex physical problems, but also have short-comings. In the final section, we will have a look at a contemporary problem in the context of galaxies where we are currently\footnote{The section was first written during the fall of 2022, but updated continuously during the winter and spring of 2023, thus highlighting the problem of keeping up with the timescales of the progress of science.} witnessing an apparent problem with the timescales of structure formation.

\section{The first galaxies}
\label{sec:jwst}

\begin{figure*}[!t]
    \centering
\minipage[t!]{\dimexpr0.98\linewidth-2\fboxsep-2\fboxrule\relax}
\begin{bclogo}[
    couleur=gray!20,
    epBord=1,
    arrondi=0.1,
    logo=\bcinfo,
    marge=8,
    ombre=false, 
    couleurBord=gray!60,
    barre=line]
    { \ \textsf{Intermezzo: The Great Debate}}
    \small{\textsf{Introductory courses on astronomy and cosmology like to begin with the story of a verbal, but heated, fight known as The Great Debate that took place on 26 April 1920 between two astronomers:
    The young, ambitious Harlow Shapley
    (who claimed he only became an astronomer because he couldn't pronounce the word ``archaeology''; \citealt{Ferris1977})
    argued that the numerous nebulous objects observed scattered across the sky were all inside our Milky Way, and that the Milky Way comprised the whole Universe.
    In contrast the older, more experienced speaker, Heber Curtis, argued that he had proof that several of these nebulae must be located much farther away, and that the Milky Way was just one of several galaxies.\vspace{1mm}\\
    The idea for this public event had been conceived between the two Solar astronomers, Charles Abbot and George Hale, who wanted a bit of diversion.
    They had also been considering a discussion about Einstein's new theory of relativity but, as Abbot wrote in letters to Hale, ``\emph{From the way the English are rushing relativity in \emph{[the journal]} Nature and elsewhere it looks as if the subject would be done to death long before the meeting\ldots}'' Abbot feared that only half a dozen in the audience would understand anything, and, as he wrote a few weeks later ``\emph{I pray to God that the progress of science will send relativity to some region of space beyond the fourth dimension, from whence it may never return to plague us}'' \citep[Archives of the National Academy of Sciences, quoted in][]{Berendzen1970}.
    Other ideas about glacial periods and zoology were also discarded, so since Abott had to be so over-dramatic they ended up agreeing on Shapley and Curtis, who were each paid \$150 \citep[out of which they had to pay their own travel expenses;][]{Trimble1995}.\vspace{1mm}\\
    One of Shapley's arguments was that, if the nebulae were separate galaxies of size comparable to the Milky Way, they should lie not just outside the Milky Way, but at unfathomable distances.
    According to measurements at the time, the distance to e.g. Andromeda would be between $\nicefrac{1}{2}$ and 100 million lightyears away; a distance which was unheard of.
    But in some of the nebulae, astronomers had observed what was thought to be \emph{novae}, transient increases in the brightness of stars.
    If the island universes were really at such enormous distances ``\emph{it would be necessary to assign what seem impossibly great absolute magnitudes to the novae}''\footnote{The Great Debate is transcribed, ostensibly, in \citet{Shapley1921}, but even without comments and discussion this text would take well over two hours to deliver, and the speakers only had 40 minutes each, so the authors must have embellished the story a bit. \citet{Hoskin1976} delves deeper into what actually happened that Monday morning in April.}.
    Moreover, the renowned astronomer Adriaan van Maanen had, four years earlier, measured the rotation of some of the spiral nebulae to be around $10^5$ years \citep{VanMaanen1916}; at extragalactic distances that would imply faster-than-light speeds in the outskirts of the disks.\vspace{1mm}\\
    On the other hand, Curtis argued that spectra of the spiral nebula did not resemble other, nearby nebulae.
    Rather, they looked like a mixture of various stars.
    Furthermore, more novae had been detected inside the Andromeda nebula than in all of Milky Way.
    If Andromeda were really inside the Milky Way, why would it host such a surplus of novae?\vspace{1mm}\\
    Unbeknownst at the time, what was thought to be novae was in fact \emph{super}novae, up to $\sim10^6$ times brighter.
    Meanwhile, van Maanen's measurements turned out simply to be erroneous.
    On the other hand, Shapley was right in several other points --- for instance his estimate of the size of the Milky Way, and the position of the Sun in it, was more correct that Curtis' --- so the Debate did not have a clear ``winner''.
    Most presumably favored Curtis, possibly because of his eloquence \citep{Ferris1977}, but Curtis regarded himself as having won the Debate \citep{Horvath2020}.\vspace{1mm}\\
    Nevertheless, the dispute was settled less than five years later when it was announced that Edwin Hubble had discovered a so-called \emph{Cepheid star} in Andromeda \citep{Hubble1925}.
    With the relation between pulsation period and absolute brightness discovered recently by Henrietta Swan Leavitt (\citeyear{Leavitt1908}), Hubble could show that Andromeda was far outside the Milky Way, sweeping all doubt away: Curtis was right, and Shapley was wrong.
    }}
\label{info:debate}
\end{bclogo}
     \endminipage
\end{figure*}

On Christmas day 2021, we finally saw the successful launch of the long-awaited James Webb Space Telescope (JWST).
Half a year later, after a month-long journey to its ``destination'' Lagrange point 2 (L2) 1.5 million kilometers from Earth, cooling to 6--40 K, adjusting of its 18 mirror segments, and commissioning and calibrating of its four instruments, the astronomy community finally saw the release of the telescope's first science data in July 2022.
A proclaimed goal of JWST is to look for some of the very first galaxies, and it was therefore expected that we would eventually find galaxies with redshifts higher than the previous ``record-holder'', GN-z11, which has a redshift of $z\sim11$ \citep{Oesch2016}\footnote{\citet{Oesch2016} measured GN-z11's redshift from the Lyman $\alpha$ break (see info box~\ref{info:lyabreak}) to $11.09_{-0.12}^{+0.08}$. A very recent study by \citet{Bunker2023} report the detection of several emission lines with JWST, deducing a redshift of $10.6034\pm0.0013$.} and is hence seen some 400 Myr after the Big Bang.

\begin{figure}[!t]
    \centering
    \includegraphics [width=0.48\textwidth] {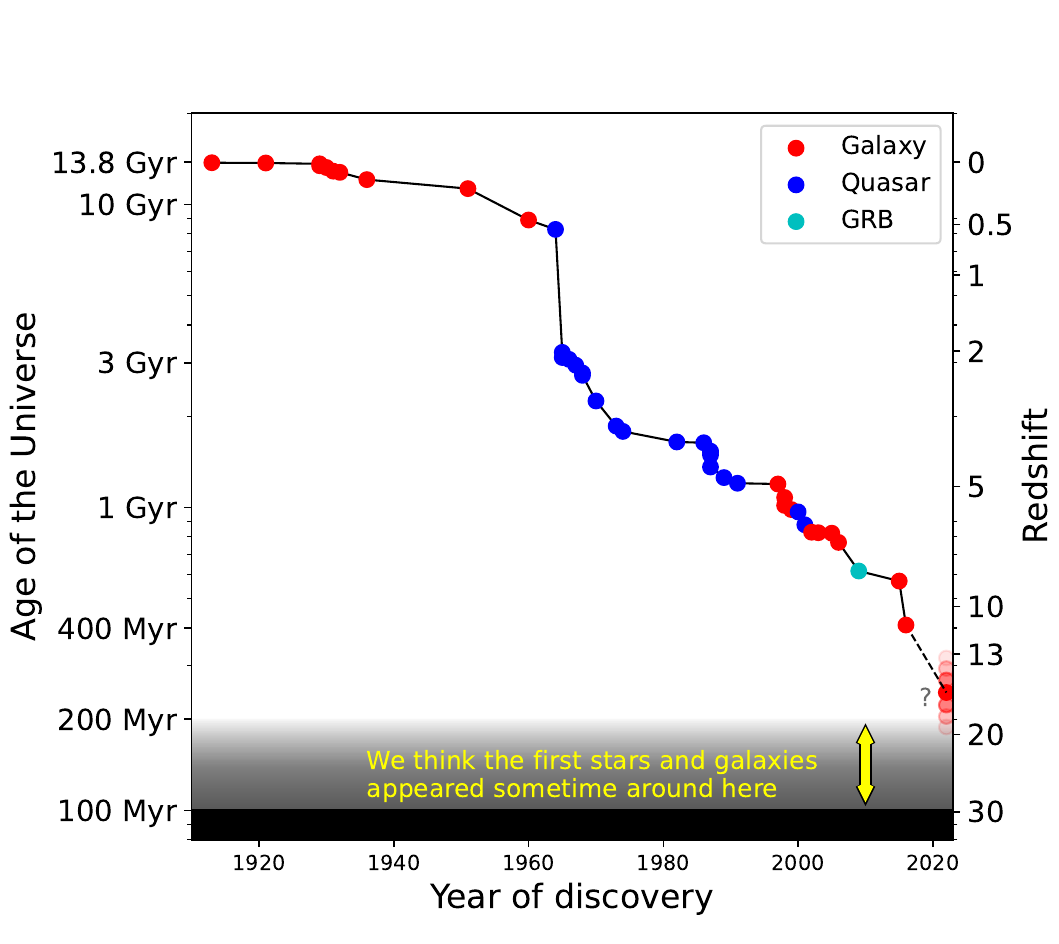}
    \caption{Earliest epoch at which an object has been discovered (and the correspinding redshift of these objects on the right $y$ axis) vs.~the year of discovery.
    Most objects are galaxies (red), but from the mid-1960s to the mid-1990s, quasars (blue) were more easily found. The cyan point indicates the gamma-ray burst GRB 090423 \citep{Tanvir2009}.
    Before JWST, the record was set by GN-z11 at $z\sim11$ \citep{Oesch2016}. 
    The gray extension indicates that this record is currently being challenged by reports of galaxies at $z=13\text{--}20$, approaching closely the epoch of the advent of the very first luminous sources at $z\gtrsim20$.}
    \label{fig:zrecord}
\end{figure}

Finding the highest-redshift galaxy is not just a matter of breaking a record. The first luminous sources are thought to appear at a redshift of $z\simeq20\text{--}30$, when the Universe was 100--200 Myr old.
Although 100 Myr is not much, cosmologically speaking, finding a galaxy 300 Myr after the Big Bang is significantly closer to the epoch of first galaxies than 400 Myr.
Figure~\ref{fig:zrecord} shows how our ``explored'' Universe has expanded, in terms of the most distant source detected.
The first galaxy to have its redshift measured (not counting the blueshifted Andromeda) was the Sombrero galaxy in 1913 with $z=0.004$ \citep{Slipher1915,Hoyt1980}. Progress was rather slow until the mid-1960s when it was realized that some bright point sources, first discovered as radio sources, were in fact extremely distant quasars \citep{Schmidt1964,Schmidt1965}
Because of the long timescales required to build up the supermassive black holes needed for a quasar to form
(but see discussion later in this section)
these objects become increasingly rare as we go to higher redshifts, and above $z\sim6$ astronomers had to wait for new techniques to discover high-redshift galaxies.
Rescue came with the so-called dropout technique \citep{Steidel1996} which since the mid-1990s has resulted in a steadily increasing highest redshift.

No more than five days after the science data started flowing from JWST, the first reports of a redshift record-breaking galaxy came when \citet{Naidu2022} announced, on the preprint server arXiv, the discovery of two galaxies with apparent redshifts of $z\simeq11$ and $z\simeq13$ (later adjusted to $z\simeq10$ and $z\simeq12$).
Simultaneously\footnote{In fact uploaded to arXiv a mere 2:02 minutes after the Naidu paper.} with this, \citet{Castellano2022} reported a sample of galaxies with redshifts up to $z\simeq15$.
In the days to follow, more astounding discoveries came, with galaxies at
$z\sim17$ \citep{Donnan2023,Harikane2022}, and even
$z\sim20$ \citep{Yan2023}.
Could this really be some of the very first galaxies, consisting of stars made of pristine gas?

More remarkably, however, is the fact that several of the galaxies seemingly are exceptionally massive; indeed the inferred stellar masses are much higher than expected to be physically possible from our understanding of hierarchical clustering and galaxy assembly.
Some even have masses comparable to the Milky Way, but ostensibly assembled in 1/20 of the timescale!

\begin{figure}[!t]
    \centering
    \includegraphics [width=0.48\textwidth] {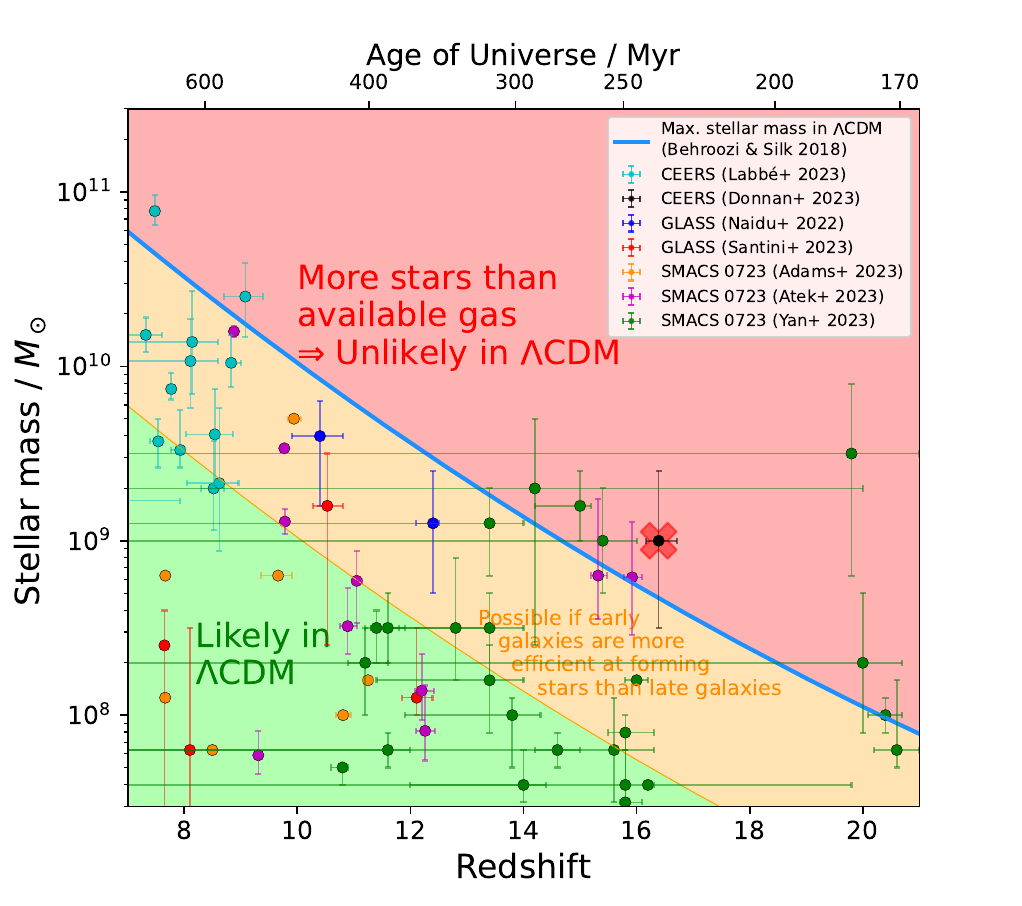}
    \caption{Stellar masses vs.~redshift of galaxies recently observed with JWST, with data taken from
    \citet{Labbe2023},
    \citet{Donnan2023},
    \citet{Naidu2022},
    \citet{Santini2023},
    \citet{Adams2023},
    \citet{Atek2023},
    \citet{Yan2023}.
    The secondary $x$ axis shows the corresponding age of the Universe at the time we see them.
    The solid blue line shows the threshold stellar masses for a cumulative number density of 1 object in a volume of $10^5\,\mathrm{Mpc}^3$, roughly equal to the volumes in which the data were obtained.
    If a survey finds with significant confidence that galaxies with stellar masses larger than this had a cumulative number density higher than $10^{-5}\,\mathrm{Mpc}^{-3}$, it would rule out $\Lambda$CDM.
    The line assumes a 100\% star formation efficiency, i.e.~with $M_*/M_\mathrm{halo} = f_\mathrm{b} \simeq 0.16$.
    Assuming a more modest star formation efficiency of 10\% --- which is still high compared to most lower-redshift models --- shifts down the line, leaving even more galaxies in tension with $\Lambda$CDM.
    The red cross marks the galaxy CEERS-93316 which was first announced as a $z\sim17$ galaxy but recently confirmed to be a member of a $z=4.9$ cluster.}
    \label{fig:jwst_stellar_mass_vs_redshift}
\end{figure}

Figure~\ref{fig:jwst_stellar_mass_vs_redshift} shows many of the newly discovered sources for which redshift and stellar masses have been determined.
The galaxies were found in three different fields of comparable survey volumes, $\sim10^5\,\mathrm{Mpc}^3$.
One is the very first image released to the public, the galaxy cluster SMACS 0723, and the two others are the Early Release Science program GLASS \citep{Treu2022} and CEERS \citep{Finkelstein2023}.
The blue line, adopted from \citet{Behroozi2018}, shows the maximum stellar mass of a galaxy expected to be found in such a volume if we assume that
1) the concordance model of the structure and evolution of the Universe, the so-called $\Lambda$CDM model, is correct, and
2) galaxies at that time were somehow able to convert 100\% of their gas to stars.
Because of the timescales needed to build up stellar mass, galaxies with masses above this line, in the red domain, are so rare that they are highly unlikely to be found in the surveyed volume.

Disturbingly, a few of the galaxies seemingly defy the timescales of $\Lambda$CDM.
The star formation efficiency is defined as the fraction of available gas that a galaxy is able to convert to stars on a dynamical timescale.
Most models require a very low efficiency to match observations, and a 100\% star formation efficiency is a factor of at least 10 more efficient than in the more nearby Universe \citep[see e.g.][]{Behroozi2013,Tacchella2018}.
If instead we assumed 10\%, the division between the two domains would move down, indicated by the intermediate, yellow region.
This would mean that even more galaxies would seem to be in tension with the timescales of structure formation.

Before claiming that not only the redshift record, but also $\Lambda$CDM has been broken, there are however a number of caveats that must be considered.
First of all, at the time of writing this text, several of the papers have not yet been peer-reviewed.
More importantly, the method by which the redshifts have been measured is a somewhat unreliable way.
Few of them have spectra, which would be needed to measure exact redshifts.
Instead the redshifts are estimated through the ``dropout technique'': At these high redshifts, the Universe is predominantly neutral (we are only in the beginning of the epoch of reionization), so all light blueward of the hydrogen ionization threshold at $912\,{\AA}$ is absorbed.
Even somewhat less energetic photons are eventually redshifted cosmologically up to the Lyman $\alpha$ wavelength at $1216\,{\AA}$ and scattered out of the line of sight \citep{Laursen2011}, and hence galaxies will be invisible in --- or \emph{drop out} of --- images taken with filters blueward of $1216\,{\AA}$.
This spectral feature where the intensity drops sharply at 1216 {\AA} is called the ``Lyman $\alpha$ break'' (se info box~\ref{info:lyabreak}).
The higher the redshift a galaxy has, the redder the filter it will drop out of, so imaging a field with multiple filters and subsequently fitting a synthetic spectrum give a (somewhat crude) estimate of the redshift.
The synthetic spectrum is calculated assuming various parameters such as stellar mass, age of the stellar population, and metallicity.
The technique therefore also provides constraints on these parameters, but one important caveat is that the parameters, to some extent, are degenerate.

\begin{figure}[!t]
\minipage[t!]{\dimexpr0.98\linewidth-2\fboxsep-2\fboxrule\relax}
\begin{bclogo}[
    couleur=gray!20,
    epBord=1,
    arrondi=0.1,
    logo=\bcinfo,
    marge=8,
    ombre=false, 
    couleurBord=gray!60,
    barre=line]
    { \ \textsf{The Lyman $\alpha$ break}}
    \small{\textsf{``Lyman $\alpha$'', or simply Ly$\alpha$, is the moniker given to light with the specific wavelength 1216 {\AA}, originating from the 2P$\rightarrow$1S transition of neutral hydrogen.
    Because a Ly$\alpha$ photon also has the exact required energy to \emph{excite} neutral hydrogen (its energy is ``in resonance'' with the 1S$\rightarrow$2P transition), because the cross section of this transition is rather large, and because neutral hydrogen is ubiquitous in galaxies, the mean free path of Ly$\alpha$ is usually small. It is therefore absorbed and instantly re-emitted in another direction, in this way \emph{scattering} its way out of the galaxies.\footnote{In fact, the energy does not have to be in exact resonance, and the scattered photon is not emitted exactly instantly: the allowed width $\Delta E$ in required energy is related to the timescale $\Delta t$ of re-emission through Heisenberg's uncertainty principle, $\Delta E \Delta t \gtrsim \hbar/2$.}\vspace{1mm}\\
    Once a Ly$\alpha$ photon escapes a galaxy, it is redshifted by the expansion of the Universe to the point that it is no longer a Ly$\alpha$ photon, after which it may travel more or less freely across the Universe toward our telescopes.
    On the other hand, the part of the galaxy's spectrum blueward of 1216 {\AA} is eventually redshifted to 1216 {\AA}, and if there happens to be some neutral hydrogen around at this point, it will be scattered out of the line of sight.\vspace{1mm}\\
    In the present-day Universe, the intergalactic medium is ionized and the full spectrum of a galaxy is transmitted.
    As we observe galaxies at increasingly large distances, we see more and more absorption lines in galaxy spectra, and as we approach the epoch of reionization these lines eventually overlap until, at $z\gtrsim6$ the spectrum is completely erased blueward of the Ly$\alpha$ line.
    Light emitted from galaxies at high redshifts hence contain a sharp ``break'' at 1216 {\AA} \emph{in the rest-frame}, and from the \emph{observed} wavelength of the break, one can then deduce the redshift of the galaxy.\vspace{1mm}\\
    Because of complex radiative transfer effects, the measured redshift is imprecise and can be confused with other spectral features if the signal-to-noise is not good.
    For instance, each scattering induces a Doppler shift to Ly$\alpha$ photons, gas bulk velocities such as galactic outflows shifts the emitted line, ionized bubbles around galaxies increases transmission, dust may preferentially absorb the wings of the emission line, and so on.
    For an introduction to these effects and how to deal with them, especially numerically, see e.g. \citet{Laursen2010}.
    }}
\label{info:lyabreak}
\end{bclogo}
     \endminipage
\end{figure}

Not only is the redshift imprecise; without a spectroscopical redshift to confirm spectral features, we cannot be certain that the galaxy is not red for some other reason than redshift. Dusty galaxies also appear red, because dust absorbs more easily shorter wavelengths. Galaxies with old stellar populations, where the massive blue stars have died, similarly appear red. Even brown dwarfs in the Milky Way may sometimes allude the observer.
For this reason the galaxies must be followed up spectroscopically.

Many other explanations have also been proposed for the too-bright galaxies.
\citet{Mason2023} show that stochastic star formation, where stars sometimes form slowly and other times fast, as well as more dust-free environments, could make a fraction of the galaxies appear much brighter than average, and that we are merely seeing ``the tip of the iceberg''.
Since the stellar mass threshold in Fig.~\ref{fig:jwst_stellar_mass_vs_redshift} is based on the average number density of galaxies, another possibility is that we have simply been lucky and pointed the telescope in a fortunate direction with galactic overdensities.

Perhaps the reason could be that the models by which we interpret the new results rely too much on physics known from the local Universe, which in fact we don't know holds true in these hitherto unexplored epochs?
\citet{Steinhardt2022} point out that the templates that are used to calculate redshifts from the series of images through different filters are based on low-redshift physics.
But there are several reasons to believe that this is not appropriate at very high redshifts.
For instance they argue that, because the Universe was hotter in the past, the mass distribution of stellar populations was significantly different from today.
As demonstrated by \citet{Sneppen2022}, accounting for the different gas temperature in the early Universe, the calculated masses of the galaxies seems to be significantly smaller, probably at least ten times.
In other words, the galaxies in Fig.~\ref{fig:jwst_stellar_mass_vs_redshift} would move down at least by 1 dex, placing most of them in the more likely (green) domain.
On the other hand, \citet{Gimenez-Arteaga2022} show that stellar masses of poorly resolved galaxies could in fact by \emph{under}-estimated by as much as 1 dex, because an older, unresolved stellar population may be outshone by young stars and hence go undetected.
This would tend to move the points up again.

Additionally, recent studies \citep[e.g.][]{Kocevski2023,Larson2023} hint at yet another possible alternative: Stellar masses are estimated from the amount of light, but now more and more high-redshift galaxies turn out to contain active galactic nuclei, the bright light of which may dominate over starlight. However, this is no less interesting; although the timescales of galaxy formation may be ``saved'', it raises the question of how supermassive black holes may form on such short timescales.
More possible explanations are discussed in recent papers such as \citet{Kannan2022}.

Even prior to JWST, some of the highest-redshift galaxies were in tension with the current paradigm, described by \citet{Steinhardt2016} as the ``impossibly early galaxy problem''.
But although at first glance the observations from JWST seem difficult to reconcile with the timescales of structure formation, there are still many pitfalls to consider before claiming the downfall of $\Lambda$CDM.

\section{Conclusion}
\label{sec:conclusion}

Modeling the multiplicity of timescales in astrophysical systems is challenged by the immense span; from sub-second processes to gigayears of evolution.
Nevertheless, this span can in some cases be an advantage in the sense that one may simply compare characteristic timescales involved in the processes to predict their outcome.
In this chapter, we have reviewed the physics of galaxy formation and seen examples of how timescales regulate which dark matter halos are able to form galaxies, and how efficient these galaxies are at forming stars.

As a more contemporary topic, we have discussed the recent reports of seemingly very early and very massive galaxies, and how they are in tension with the timescales of early structure formation.
The $\Lambda$CDM model has been extremely successful not only as an explanation of existing observations, but also in making predictions such as weak gravitational lensing \citep{Fischer2000}, polarization of the cosmic microwave background \citep{Kovac2002}, and patterns in the large-scale structure of the matter distribution
\citep{Eisenstein2005}.
However, the model builds upon three somewhat unstable pillars, namely 1) general relativity which despite its success on large scales is yet to be unified with quantum mechanics, and 2) dark matter and dark energy, the nature of which is unknown and even questioned by a significant fraction of the scientific community.
While it is thus not at all unlikely that we will at some point have to face a cosmological paradigm shift, there are many reasons to wait making too strong statements about $\Lambda$CDM; we will have to await more data, in particular spectroscopic data and more wide-field surveys.
Equally importantly, we should start thinking seriously about expanding the low-$z$ models that are currently being applied, probably unfairly, to interpret high-$z$ data.

\begin{center}
\pgfornament[width=.4\textwidth]{89}
\end{center}

\begin{acknowledgements}
    I am most thankful to the referee for numerous excellent comments and suggestions, as well as to Charles L.~Steinhardt, Gabriel Brammer, and Steen H.~Hansen for fruitful discussions.
    The Cosmic Dawn Center (DAWN) is funded by the Danish National Research Foundation under grant No. 140.
\end{acknowledgements}

\bibliographystyle{aa}
\bibliography{timescales}

\end{document}